\documentclass[%
 reprint,
 longbibliography,
 superscriptaddress,
groupedaddress,
 amsmath,
 amssymb,
 aps,
 prx,
 floatfix,
]{revtex4-2}
\usepackage[utf8]{inputenc}
\usepackage{graphicx}
\usepackage{dcolumn}
\usepackage{bm}
\usepackage[dvipsnames]{xcolor}
\usepackage[colorlinks=true,linkcolor=teal,citecolor=Maroon,urlcolor=Maroon]{hyperref}
\usepackage{physics}
\usepackage{siunitx}
\usepackage{float}
\usepackage{tikz}
\usetikzlibrary{quantikz}
\usepackage{mathbbol}
\usepackage{enumitem}
\usepackage{xfrac}

\newcommand{\secref}[1]{Section~\hyperref[#1]{\ref*{#1}}}
\newcommand{\appref}[1]{Appendix~\hyperref[#1]{\ref*{#1}}}
\newcommand{\tabref}[1]{Table~\hyperref[#1]{\ref*{#1}}}
\newcommand{\figref}[1]{Fig.~\hyperref[#1]{\ref*{#1}}}
\newcommand{\sfigref}[2]{Fig.~\hyperref[#1]{\ref*{#1}(#2)}}

\usepackage{svg}
\newcommand{\orcid}[1]{\href{https://orcid.org/#1}{\includegraphics[width=8pt]{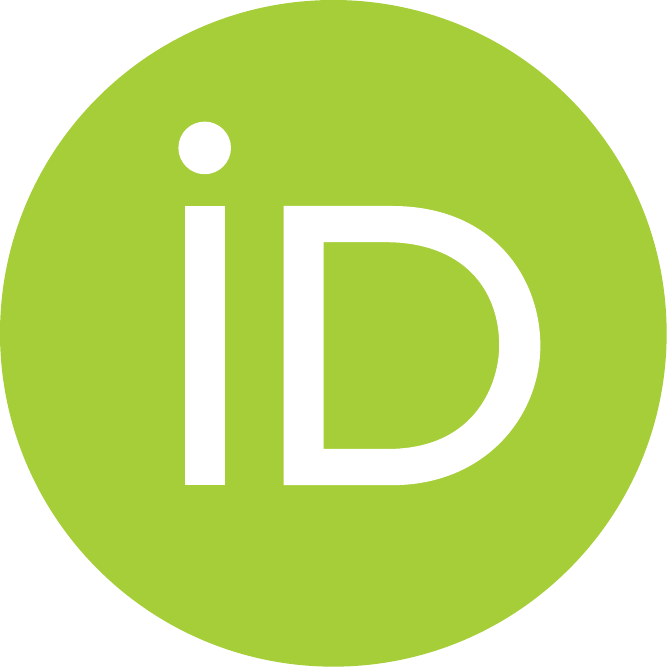}}}

\begin{document}

\title{Autonomous coherence protection of a two-level system in a fluctuating environment}
\author{Fernando Quijandría\orcid{0000-0002-2355-0449}}
\email{isaac.quijandria@oist.jp}
\author{Jason Twamley\orcid{0000-0002-8930-6131}}
\affiliation{Quantum Machines Unit, Okinawa Institute of Science and Technology Graduate University, Okinawa 904-0495, Japan}

\date{\today}
\begin{abstract}
We re-examine a scheme generalized by [R. Finkelstein et al, Phys. Rev. X 11, 011008 (2021)], whose original purpose was to remove the effects of static Doppler broadening from an ensemble of non-interacting two-level systems (qubits). 
This scheme involves the simultaneous application of red and blue detuned drives between a qubit level and an auxiliary level, and by carefully choosing the drive amplitudes and detunings, the drive-induced energy shifts can exactly compensate the inhomogenous static Doppler-induced frequency shifts - effectively removing the inhomogeneous Doppler broadening.
We demonstrate that this scheme is far more powerful and can also protect a single (or even an ensemble), qubit's energy levels from noise which depends on both time and space: the same scheme can greatly reduce the effects of  dephasing noise induced by a time-fluctuating environment. As examples we study protection against two types of non-Markovian environments that appear in many physical systems: Gaussian noise and non-Gaussian noise - Random Telegraph Noise.  Through numerical simulations we 
demonstrate the enhancement of the spin coherence time $T_2^*$, of a qubit in a fluctuating environment by three orders of magnitude as well as the refocusing of its initially drifting frequency. This same scheme, using only two drives, can operate on an collection of qubits, providing temporal and spatial stabilization  simultaneously and in parallel yielding a collection of high quality near-identical qubits which can be useful for many quantum technologies such as quantum computing and sensing, with the potential to achieve fault tolerant quantum computation much sooner. 
\end{abstract}

\maketitle

\section{Introduction\label{sec:introduction}}

Quantum systems such as qubits unavoidably interact with their surrounding environment. This interaction leads to the loss of quantum coherence which is detrimental for any quantum technology. Perhaps the two best-known strategies to deal with decoherence are Quantum Error Correction (QEC) and Dynamical Decoupling (DD).
Conventional QEC requires the encoding of a logical qubit into several physical qubits and permits one to detect and correct errors without learning the state of the logical qubit~\cite{Nielsen2010QuantumInformation}. An alternative approach uses the infinite Hilbert space of a harmonic oscillator to redundantly encode quantum information~\cite{Joshi2021QuantumQED}. 
QEC is well-suited to deal with Markovian environments as errors occurring in the system can always be described as completely-positive maps~\cite{Terhal2015QuantumMemories}.
 However, in many realistic situations the noise cannot be approximated as Markovian and environmental correlations must be taken into account. 

DD corresponds to control methods inspired by the Hahn echo pulse sequence~\cite{Hahn1950SpinEchoes}. 
It can be classified into discrete dynamical decoupling (DDD) which relies on discrete pulse sequences that, when acting on time scales smaller than the correlation time of the environment, can revert and therefore mitigate its effects on the system, 
and continuous dynamical decoupling (CDD)  which uses continuous wave drives instead of pulse sequences.
DDD sequences have been successfully realized in a broad range of quantum technologies in order to protect the coherence of a single qubit and ensembles of them from a slowly fluctuating environment~\cite{Koppens2008SpinDot, Biercuk2009OptimizedMemory, Damodarakurup2009ExperimentalControl, Du2009PreservingDecoupling, deLange2010UniversalBath, Bylander2011NoiseQubit, Peng2011HighExperiment, Bar-Gill2013Solid-stateSecond, Ezzell2022DynamicalSurvey}.
CDD protocols are particularly attractive owing to their simplicity as compared to the more experimentally challenging to implement DDD sequences.
Similarly to their discrete counterpart, 
different CDD protocols addressing a single and ensembles of qubits, have been proposed and realized experimentally in different physical systems ~\cite{Bermudez2012RobustDecoupling, Cai2012RobustDriving,Golter2014ProtectingStates,Cohen2017ContinuousDetuning,Teissier2017HybridSpin, Farfurnik2017ExperimentalDecoupling, Aharon2019RobustDecoupling, Genov2019MixedDecoupling, Wang2020CoherenceDriving}. Recently, the numerical optimization of some of the above protocols has also been studied~\cite{Cai2022OptimizingLearning}.

The mitigation of noise in quantum system, and qubits in particular, is a powerful tool to improve the performance of the physical qubit layer within QEC schemes. In particular the resources required to perform {\it fault tolerant} QEC will be greatly reduced if the error rates of the physical qubits can be pushed to be far below the fault tolerant thresholds.

In this work we demonstrate the autonomous correction (mitigation), of the stochastic fluctuations in the frequency of a qubit resulting from its coupling to a fluctuating non-Markovian environment. 
This is achieved by the continuous driving of a transition between the qubit and an auxiliary energy level which is more sensitive to the same source of noise. 
In some sense, the auxiliary level  
permits the system to gain access to the noise's temporal behavior. 
The latter is imprinted in the drive-induced energy shifts in the qubit transition which can exactly compensate the stochastic fluctuations by carefully choosing the amplitudes and the detunings of the driving fields.
The studied protocol was originally introduced in  Ref.~\cite{Yavuz2013SuppressionShift} and later generalized in Ref.~\cite{Finkelstein2021ContinuousDephasing} to deal with the Doppler broadening of atomic transitions. Whereas these works considered only time-independent frequency fluctuations, here we demonstrate that the same protocol is able to deal with time-dependent noise.
We  study the cases of Gaussian and non-Gaussian stochastic non-Markovian noise and we characterize how well this scheme can protect the qubit's coherence (as an increase of the dephasing time), as a function of the correlation time of the noise and other parameters of the protocol.  
Finally, we demonstrate that this continuous wave protocol is compatible with the application of fast single-qubit gates. As examples we  consider the cases of a single-qubit rotation and a Hadamard gate. 

\section{Continuous off-resonance protection from static inhomogeneous dephasing using ac-Stark shifts revisited}

\begin{figure}
\centering
\includegraphics[width=1.0\linewidth]{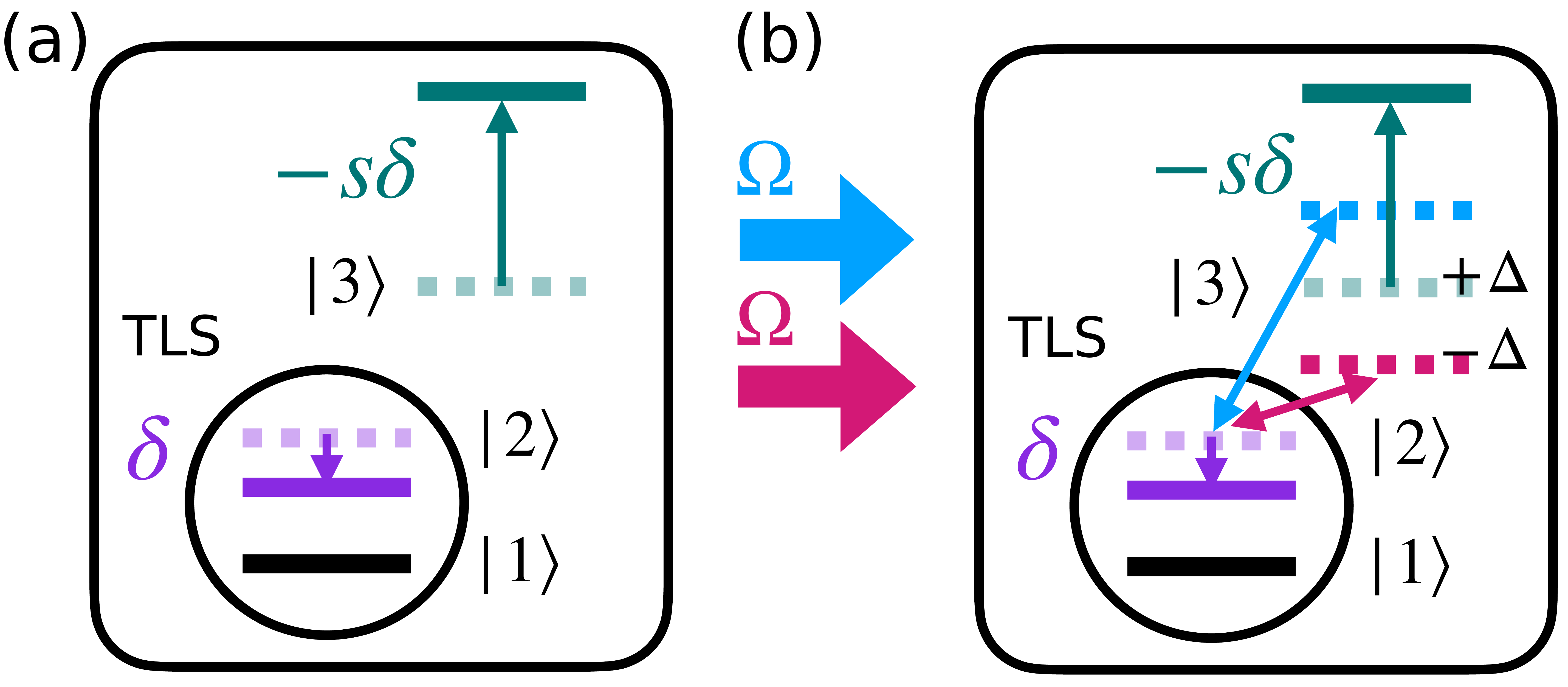}
\caption{Sketch of the protection scheme against inhomogeneous fluctuations of a TLS transition frequency. (a) A TLS is formed by the two lowest energy levels $\{ \ket{1},\, \ket{2} \}$ of a multilevel quantum system - the emitter. 
Due to the presence of the environment the different energy transitions of the emitter 
are shifted.  
We represent this by the shifted energy levels $\ket{2}$ (purple) and $\ket{3}$ (teal) in solid lines with respect to their homogeneous values in dashed lines. 
We consider a system in which these fluctuations are not independent but are given by the stochastic 
parameter
$\delta$ and $-s \delta$ respectively, with $s \geq 1$ .
(b) The protection of the TLS transition frequency from the 
inhomogeneous
fluctuations is achieved by driving 
the transition between the excited TLS level $\ket{2}$ and the auxiliary level $\ket{3}$ with two
continuous-wave drives of the same amplitude $\Omega$ red- and blue-detuned by $\Delta$ from the above mentioned transition.}
\label{fig:sketch}
\end{figure}

We first introduce the protocol as it was originally devised to combat static spatial inhomogeneous broadening. The authors in Refs.~\cite{Yavuz2013SuppressionShift, Finkelstein2021ContinuousDephasing} studied how to mitigate  dephasing noise which results from static inhomogenous variations  in the transition frequencies of an ensemble of \textit{non-interacting} two-level systems using the quantum atom-optical process known as the ac-Stark shift or light shift. By this we refer to the shift in the energy levels of a two-level system due to its continuous driving by an off-resonant classical field. Before revisiting their protection scheme, let us settle down the notation. The goal in the above references, and our work as well, is to protect a two-level system (TLS) or qubit from the decoherence induced by an inhomogeneous environment. 
We consider that the TLS is comprised of two energy levels which are contained within a more general multilevel system, the latter  we  call henceforth as the emitter. 

In order to autonomously correct for time-independent frequency variations in a TLS ensemble we additionally include a single, or a pair of, auxiliary energy levels which are also sensitive to the same source of inhomogeneity affecting the TLS ensemble. From here on we are going to focus on the scheme in Ref.~\cite{Finkelstein2021ContinuousDephasing} which makes use of a single auxiliary energy level. Therefore, let us consider now an ensemble of three-level systems $\{ \ket{1}, \ket{2}, \ket{3} \}$.
In the absence of any inhomogeneity each emitter in the ensemble has identical energies for these three levels. We consider the TLS to be made from $\{ \ket{1}, \ket{2}\}$, while level $\ket{3}$ is the auxiliary level. 
We will assume the homogeneous energy of the ground state $\ket{1}$ to vanish while the homogeneous energies of the levels $\ket{2}$ and $\ket{3}$ to be $\omega_2$ and $\omega_3$ respectively.
In the presence of the inhomogeneous (but static), environment, the Hamiltonian of the emitter ensemble is
\begin{align}
    H &= \sum_i (\omega_2 - \delta_i)  \ket{2} \bra{2} + (\omega_3 + s \delta_i) \ket{3} \bra{3},
\end{align}
where $\delta_i$ is a random number drawn from a distribution with width $\sigma$ which characterizes the inhomogeneous shift in the frequency of the emitter. 
Notice that we have introduced the scaling parameter $s \geq 1$ and that, by definition, the inhomogeneous frequency shift of level $\ket{3}$ is $-s$ times that of level $\ket{2}$. This is depicted in Fig.~\ref{fig:sketch}(a).

We now  introduce two driving fields, each with amplitude $\Omega$ but opposite detunings $\pm \Delta$ (for $\Delta > 0$), driving the transition $\ket{2}\leftrightarrow\ket{3}$. 
In a doubly-rotating frame with frequencies $\omega_2$ and $\omega_3$,
the total Hamiltonian of the $i$-th emitter is given by $H^i_{\rm rot}(t) = H^i_{\rm inh} + H_{\rm drive}(t)$ with
\begin{align}
    H^i_{{\rm inh}} &= - \delta_i  \ket{2} \bra{2} + s \delta_i \ket{3} \bra{3},\label{eq:ith-emitter}
\end{align}
the inhomogeneous Hamiltonian.
Because the emitters in the ensemble do not interact with each other we can treat them independently. 
On the other hand
\begin{align}\label{eq:drive23}
    H_{\rm drive}(t)&=  \frac{\Omega}{\sqrt{2}} \left( \ket{2} \bra{3} {\rm e}^{+i \Delta t} + {\rm H.c.}  \right) \nonumber\\
    &+ \frac{\Omega}{\sqrt{2}} \left( \ket{2} \bra{3} {\rm e}^{-i \Delta t} + {\rm H.c.}  \right) ,
\end{align}
is the Hamiltonian of the drives in the rotating wave approximation (RWA) ($\omega_2,\,\omega_3 \gg \Omega$). 
Treating the two drives independently and considering the energy hierarchy (large off-resonance regime), 
$\Delta \gg \Omega, \, s \sigma$ together with $s \gg 1$,
the total Stark shift correction to level $\ket{2}$ is given by 
\begin{align}\label{eq:E-light}
    E_{\rm Stark} &\approx - \frac{\Omega^2}{2(\Delta + s \delta_i)} + \frac{\Omega^2}{2(\Delta - s \delta_i)} \nonumber\\
    &= -\frac{\Omega^2}{2\Delta} + \frac{1}{2} \left( \frac{\Omega}{\Delta} \right)^2 s \delta_i +
    \frac{\Omega^2}{2\Delta} + \frac{1}{2} \left( \frac{\Omega}{\Delta} \right)^2 s \delta_i \nonumber\\
    &= \left( \frac{\Omega}{\Delta} \right)^2 s \delta_i .
\end{align}
In other words, the effective Hamiltonian of the $i$-th TLS becomes $H^i_{\rm eff} \approx (-\delta_i + E_{\rm Stark}) \ket{2} \bra{2}$.  
If the auxiliary level is very sensitive to the inhomogeneity ($s \gg 1$), from Eq.~\eqref{eq:E-light}, we can cancel out the inhomogeneous shift of \textit{every} TLS in the ensemble by setting $(\Omega / \Delta)^2 = 1/s$, the so-called \textit{protection condition}, so that $-\delta_i + E_{\rm Stark} \approx 0$. 
We stress that this result is valid up to third-order perturbation theory in $\delta_i/ \Delta$.
The authors in Ref.~\cite{Finkelstein2021ContinuousDephasing} have generalized the above treatment beyond the large off-resonance regime. In a general setting, the protection condition reads,
\begin{align}\label{eq:protection-general}
    J_0 \left( 2 \sqrt{2}\frac{\Omega}{\Delta} \right) = \frac{s-1}{s+1}
\end{align}
with $J_0$ the zeroth-order Bessel function.
Notice that for a fixed amplitude $\Omega$, the
smaller the sensitivity $s$ the smaller the detuning needed to compensate for the inhomogeneous shift and therefore, the more the drives dress the TLS transition. We will come back to this issue later.

\section{Time-dependent noise correction}

We have just reviewed the continuous protection scheme for an ensemble of non-interacting emitters affected by time-independent variations in their transition frequencies. 
One could instead change the paradigm and consider the case of a single emitter in a time-fluctuating environment with a finite correlation time $\tau$. Here, ensemble averages are replaced by averages over time traces. 
In this work we are going to focus on a  randomization model of dephasing instead of a full quantum mechanical treatment of the interaction between the emitter and its environment. The former model supposes that the net  effect of the environment is to induce time-dependent stochastic fluctuations of the qubit frequency 
\begin{equation}
    H_{\rm TLS}(t) = \left[ \omega_2 + \delta(t) \right] \ket{2}\bra{2} ,
\end{equation}
with $\omega_2$ the homogeneous qubit frequency and $\delta(t)$ the environment-induced fluctuations. This model describes, for example, the interaction between a central spin and a large ensemble of spins far-off resonance from it. 
In a frame rotating with $\omega_2$, the initial superposition $\ket{\Psi(0)} = \ket{1} + \ket{2}$ evolves into $\ket{\Psi(t)} = \ket{1} +\exp(-i \int_0^t {\rm d}t'\, \delta(t')) \ket{2}$ after a time $t$ due to the above Hamiltonian (where we have omitted normalization factors). Because of the stochastic nature of $\delta(t)$, the accumulated phase over a fixed time $t$ is not constant and varies over different realizations. Thus, on average, there is not a fixed phase relation between the qubit states due to the fluctuating environment. 

We now return to the Hamiltonian in Eq.~\eqref{eq:ith-emitter}, and note that the index $i$ labels the spatial inhomogeneity of the energy landscape for the non-interacting ensemble. We now posit that formally the spatial average can be replaced by a temporal average and one could reinterpret Eq.~\eqref{eq:ith-emitter}, as a single emitter undergoing temporal noise over discrete time intervals. This motivates us investigate how the protection scheme of Ref.~\cite{Finkelstein2021ContinuousDephasing}, can be reinterpreted to protect against continuous temporal dephasing noise on a single emitter.
In the following we will study numerically how the correction performs as a function of the correlation time of the bath. For this, we will consider two instances of stochastic processes describing many forms of realistic fluctuating environments. These are  Ornstein-Uhlenbeck (OU) noise and  Random Telegraph Noise (RTN). 
The protection scheme is the same as the one shown in Fig.~\ref{fig:sketch}, however, $\delta$ will now correspond to a time-dependent stochastic function, i.e., $\delta \equiv \delta(t)$.

The simulations are performed using the \texttt{qutip} package~\cite{Johansson2013QuTiPSystems}, and the data is analyzed and visualized utilizing NumPy~\cite{Harris2020ArrayNumPy} and Matplotlib~\cite{Hunter2007Matplotlib_A_2D_Graphics_Environment}. 

\subsection*{Ornstein-Uhlenbeck (OU) noise}

The Ornstein-Uhlenbeck (OU) stochastic process is a phenomenological model of the Brownian motion of a particle in a surrounding medium~\cite{Uhlenbeck1930OnMotion}. It describes a Gaussian and temporally homogeneous process. In addition, it can also provide an accurate description of the effects of a spin bath on a central spin such as the NV center in diamond. The latter is true provided that the number of spins in the bath is large enough so that we can apply the central limit theorem as well as neglect the backaction of the central spin on the bath~\cite{Hanson2008CoherentBath,deLange2010UniversalBath}. 
In our case, we will consider the OU process $\delta_{\rm OU}(t)$ with zero expectation value $\langle \delta_{\rm OU}(t) \rangle = 0$ and correlation function
\begin{align}
    \langle \delta_{\rm OU}(t) \, \delta_{\rm OU}(0) \rangle = b^2 \exp \left( -t /\tau \right),
\end{align}
to describe the fluctuations of the qubit frequency, i.e., $\delta(t) \equiv \delta_{\rm OU}(t)$. Here $\tau$ is the correlation time of the environmental fluctuations and $b$ characterizes their strength~\cite{Cai2012RobustDriving, Stark2017Narrow-bandwidthDecoupling}. 

The decoherence of a central spin due to a fluctuating bath is characterized by its free induction decay (FID) or simply, its free evolution. For the case of a Gaussian environment, the decay of the initial coherence of a qubit prepared in a superposition state lying along the equator of the Bloch sphere can be analytically calculated~\cite{deSousa2009ElectronFluctuations}. Experimentally, this decay is measured in a Ramsey-type experiment in which the qubit coherence is mapped into the expectation value of the $Z$-Pauli operator $S_z$. Thus, for a Gaussian environment we have the general result $\langle S_z(t) \rangle = \exp \left[ -b^2 \tau^2 ( {\rm e}^{-t/\tau} + t / \tau - 1 ) \right]$. The relation between $\tau$ and the average time $1/b$ in which the bath fluctuates sets the \textit{slow} and the \textit{fast} bath limits. 
In the slow bath limit: $\tau \gg 1/b$, and for this bath when $t \ll \tau$, the coherence decay reduces to $\langle S_z(t) \rangle = \exp \left( -b^2 t^2 /2 \right)$ which is characterized by the decay time $T_2^{* \rm slow} = \sqrt{2}/b$.
In the fast bath regime: $\tau < 1/b$, we have $\langle S_z(t) \rangle = \exp \left[ -(b^2 \tau) t \right]$ and correspondingly a decay time of $T_2^{* \rm fast} = 1/(b^2 \tau)$.

We numerically simulate a OU process using the exact algorithm introduced in Ref.~\cite{Gillespie1996ExactIntegral}. 
We primarily study the case of a slow bath as it is more physically relevant to many solid-state spin dephasing processes and neglect energy relaxation, which ultimate limits the coherence decay. 
We scale time in units of the bath correlation time $\tau$, and choose $b =19/ \tau$. 
Following the above relations, the coherence decay time corresponding to these parameters is $T_2^{* \rm slow} = 0.07 \tau$. 
In order to show that we can actually extend the coherence time beyond $T_2^{* \rm slow}$, we simulate a Ramsey experiment: we initialize the TLS in the ground state followed by an instantaneous rotation to the $+$ eigenstate of the $S_x$ operator. We let it evolve for a time $t$, and finally we project it to the $Z$-axis and measure the expectation value of the corresponding Pauli operator $S_z$ which gives us a measure of the TLS coherence. In the absence of any dephasing the Ramsey sequence yields a final expectation value $\langle S_z \rangle = 1$, while for an environment which completely decoheres the TLS during the time period $t$, 
the Ramsey sequence returns  $\langle S_z \rangle = 0.5$.

\begin{figure}
\centering
\includegraphics[width=1.\linewidth]{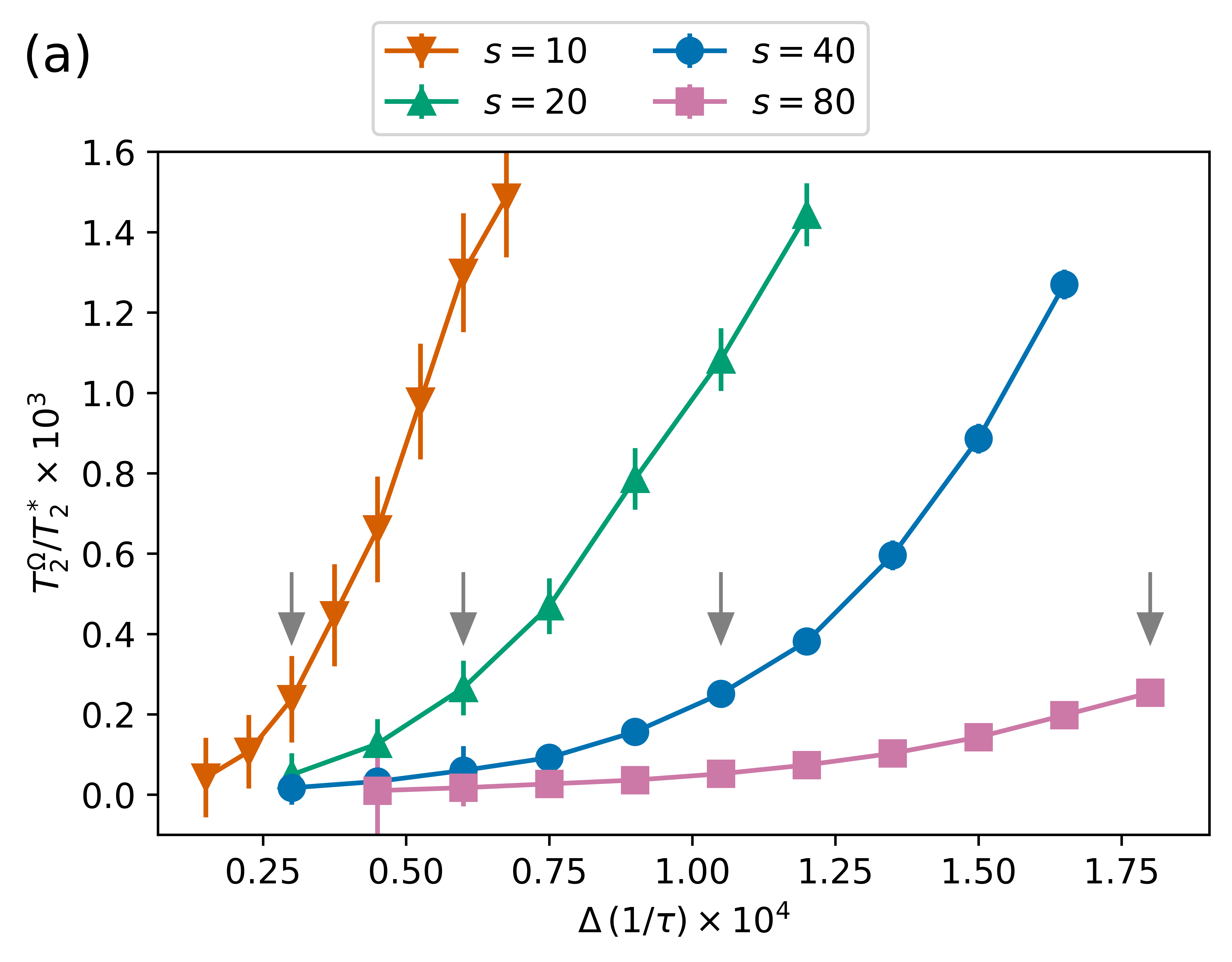}
\includegraphics[width=1.\linewidth]{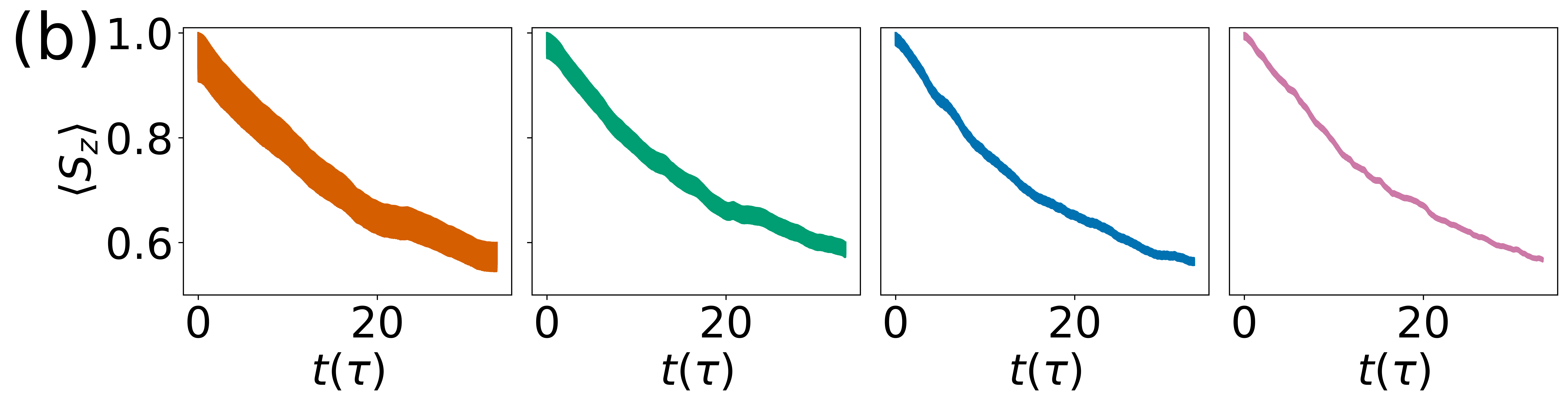}
\caption{Graphs illustrating the increase in coherence time provided by the protection scheme: (a) Coherence time gain $T_2^\Omega / T_2^*$ as a function of the detuning $\Delta$ for a doubly-driven TLS with time-dependent stochastic frequency fluctuations described by an Ornstein-Uhlenbeck process and
for different values of the sensitivity: $s=10$ (inverted triangles), $s=20$ (triangles), $s=40$ (circles) and $s=80$ (squares).
(b) Coherence decay $\langle S_z \rangle$ as a function of time for the parameter sets $(s,\Delta)$ indicated by the gray arrows in (a).}
\label{fig:ramsey}
\end{figure}

In Fig.~\ref{fig:ramsey}(a) we show the coherence time gain $T^{\Omega}_2  / T_2^{* {\rm slow}}$ of the doubly-driven TLS for different values of the sensitivity $s$ of the emitter's third level and as a function of the detuning $\Delta$ in units of $1/\tau$. In order to extract the coherence times, 
we let the driven system evolve for a time $t = 33.3 \tau$ and we calculate the expectation value of $S_z$ as a function of time. 
For every pair $(s,\Delta)$, we fix the value of $\Omega$ according to the protection condition~\eqref{eq:protection-general}. Solely satisfying~\eqref{eq:protection-general} does not guarantee the mitigation of the qubit frequency fluctuations as we also need to satisfy the energy hierarchy $\Delta \gg \Omega,\, s\sigma$ which leads to the perturbative treatment of the drive-induced energy shifts. 
For the case of time-dependent frequency fluctuations, $\sigma$ would correspond to the bandwidth of the noise. 
The OU noise exhibits a Lorentzian spectral density with a frequency cutoff proportional to $1/\tau$ which approximately determines the bandwidth.
Thus, we expect the qubit coherence to be sensitive to the frequency noise for moderate values of the detuning $\Delta$. For this reason, for every pair $(s,\Delta)$ we take averages over $10^4$ realizations of the noise. 
We fit the decay of $\langle S_z \rangle$ to an exponential decay of the form $[\exp(-t/T^\Omega_2) + 1]/2$ and
from this fitting we extract the value of the coherence decay time $T^\Omega_2$. 

As an example, in Fig.~\ref{fig:ramsey}(b) we show the full time evolution of $\langle S_z \rangle$ for four different sets of parameters $(s,\, \Delta)$. 
For each of these sets,
the corresponding full time evolution oscillates very fast in time and appears as a thick solid line.  
In order to make a fair comparison, for Fig.~\ref{fig:ramsey}(b) we have chosen parameters $(s,\, \Delta)$ which result in similar coherence decay times 
$T^\Omega_2 \approx 17.3 \tau$. These sets of parameters correspond to the gray arrows in Fig.~\ref{fig:ramsey}(a).
As we can see, for smaller values of the sensitivity $s$, $\langle S_z \rangle$ exhibits larger amplitude oscillations. This can be easily understood from the protection condition~\eqref{eq:protection-general}. 
According to it, the smaller the sensitivity $s$, the larger the value of the ratio $\Omega / \Delta$ or, in other words, the closer to resonance we need to drive the emitter. 
The observed high-frequency oscillations are a consequence that the bare TLS eigenstates, in general,  do not diagonalize the Hamiltonian of the doubly-driven emitter.
Nevertheless, as we increase the detuning of the drives we decrease the dressing of the emitter and thus the closer the dressed TLS eigenstates are to the bare TLS eigenstates. 
This is very important as the finite modulation amplitude of the fast oscillations of $\langle S_z \rangle$ will reduce the maximum average coherence attainable below unity regardless of an improved coherence decay time. 
In order to stress this, we have included  error bars in Fig.~\ref{fig:ramsey}(a). These are a measure of the amplitude of these oscillations about the mean. For these error bars to be visible on the scale of the improved coherence time, we have rescaled all of them by a factor of $\sim \cross 5$. 
This shows that while it is possible to achieve a three-order of magnitude improvement in the coherence time of the TLS 
with a relatively small detuning, 
the price to pay is a reduced maximum coherence. 
This can be further improved by using a system with a larger value of the sensitivity and accordingly larger values for the detuning,
but the maximal  degree of attainable protection  will depend on the particular physical implementation.  

In the next subsection we are going to study the case of a TLS affected by non-Gaussian noise. In particular, we are going to focus on the correction performance as a function of the correlation time of the environment. 

\begin{figure*}
    \centering   
    \includegraphics[width=180mm]{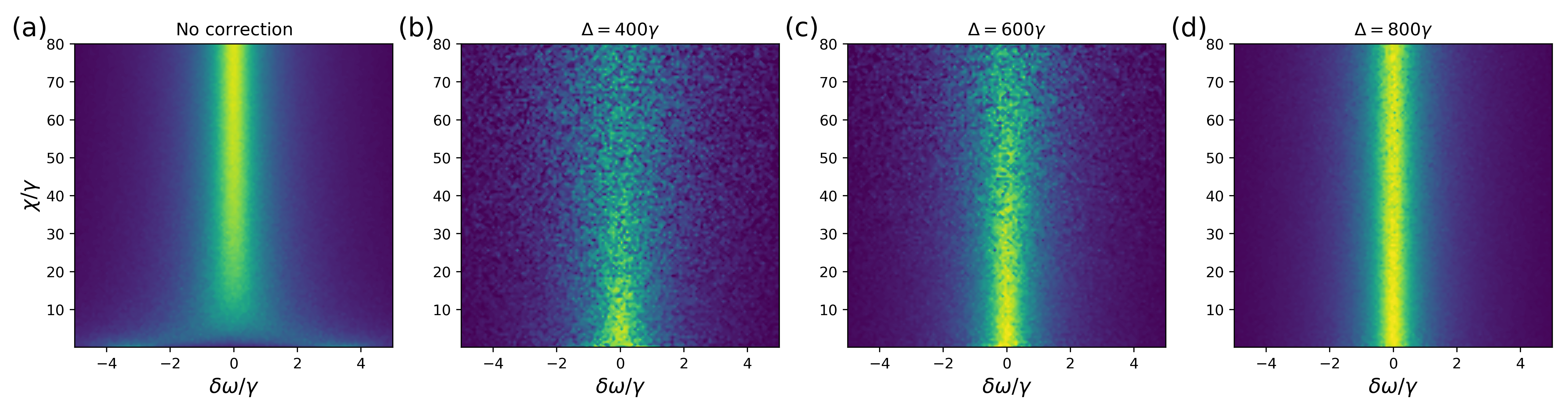}
    \caption{Response of a weakly probed TLS in a fluctuating environment modelled by RTN as a function of the detuning $\delta \omega$ of the probe drive from the TLS transition frequency, and the jump rate $\chi$ of the noise. (a) TLS response in the absence of the correction drives. We can clearly distinguish between the slow and the fast bath regimes in which the TLS responds at two frequencies and at their average value respectively. In (b)-(d) we show the response of the TLS in the presence of the correction drives for detunings $\Delta = 400 \gamma, \, 600 \gamma$ and $800 \gamma$ respectively. All of these cases correspond to an implementation with a third level with sensitivity $s=40$ and we show averages over 100 realizations of the noise. The width of the response is limited by the decay rate $\gamma$ of the TLS. We simulate the TLS relaxation using a Gorini-Kossakowski-Sudarshan-Lindblad master equation.}
    \label{fig:rtn_map}
\end{figure*}

\subsection*{Random Telegraph Noise (RTN)}

This corresponds to a non-Gaussian process in which the stochastic variable can take only one of two possible values. This model describes low-frequency fluctuations in microwave qubits and resonators~\cite{Burnett2019DecoherenceQubits,Niepce2021StabilityDefects} and has also been implemented via the modulation of a flux qubit in order to study the phenomenon of motional narrowing in a superconducting circuit~\cite{Li2013MotionalQubit}. 
In this case the TLS frequency changes randomly between the values $\pm \xi$ with jump rate $\chi$.
The probability of $n$ jumps in a time interval $t$ is given by the Poisson distribution $P_n(t) = (\chi t)^n {\rm e}^{- \chi t}/ n!$~\cite{Li2013MotionalQubit}. 

Here, instead of calculating the improved coherence time of the TLS, we are going to study the frequency response of the TLS to a weak probe signal in order to visualize the suppression of the time-dependent frequency fluctuations.
We will consider
a weak transversal probe of the form $H_{\rm probe}(t) = g \cos(\omega_p t) (\ket{1} \bra{2} + {\rm H.c.})$ with $g \ll \omega_2$, and consider as well the energy relaxation of the TLS described by the Gorini-Kossakowski-Sudarshan-Lindblad master equation $\partial_t \rho = -i [H(t) + H_{\rm probe}(t), \rho] + (\gamma/2)\mathcal{D}[\ket{1} \bra{2}] \rho$, with $\mathcal{D}[O]\rho = 2 O \rho O^\dagger - O^\dagger O \rho - \rho O^\dagger O$.
We scale time to be in units of the decay rate of the TLS $1/\gamma$.
For the rest of the parameters we choose: $\xi/\gamma = 4$, $g/\gamma = 0.1$, $s=40$ and we let the system evolve for a time $t = 15\gamma$. In Fig.~\ref{fig:rtn_map} we plot the qubit excitation probability ${\rm Tr}(\ket{2} \bra{2} \rho)$ as a function of the detuning of the probe from the TLS homogeneous frequency $\delta \omega = \omega_p - \omega_2$
and the RTN jump rate $\chi$ for (a) no correction and 
correction with an increasing detuning (b) $\Delta = 400\gamma$, (c) $\Delta = 600\gamma$ and (d) $\Delta = 800\gamma$. For (b)-(d), we have set the drive amplitude $\Omega$ to fulfil the protection condition~\eqref{eq:protection-general}. These graphs correspond to averages over 100 realizations of the noise. In this case, the coherence of the TLS is limited by the decay rate $\gamma$ which also sets the width of the TLS response (green center region).

In the absence of any correction, we can clearly distinguish between the two bath regimes: the slow one in which the jump rate is smaller than the noise amplitude $\chi < \xi$, and the fast regime in which $\chi > \xi$. In the former, we can clearly resolve two values of the frequency, i.e.,  the two 
 spots at the bottom of Fig.~\ref{fig:rtn_map}(a).
While in the fast regime the fast fluctuations quickly average out and the TLS only responds at the average frequency, that is, on resonance, i.e., the green center region in Fig.~\ref{fig:rtn_map}(a). 
In the presence of the two drives, we can see that it is possible to refocus the TLS so that it only responds on resonance.
Nevertheless, we observe that the correction refocuses the TLS up to some threshold value of the RTN jump rate $\chi$. The latter threshold seems to increase with the value of the detuning. This can be understood in the following way. First of all, the correlation function of the RTN is given by the relation $\langle \xi(t) \, \xi(0) \rangle = \xi^2 \exp \left( -2 \chi t \right)$. 
This means that the RTN power spectral density is also Lorentzian with
a natural cutoff proportional to $\chi$ (inverse correlation time of the environment). Therefore, the noise bandwidth increases linearly with $\chi$ and correspondingly we need to increase the detuning of the drives in order 
to be sufficiently detuned away from it. 

Thus, in principle, the presented correction scheme is not limited by the correlation time of the environment. The limitations to its performance will be imposed by the physical implementation of the scheme, in particular, by the achievable detunings from the TLS's auxiliary level transition. For instance, for the case of a slightly anharmonic oscillator (Kerr or Duffing oscillator), the formulas for the ac Stark shifts need to be corrected to take into account the presence of other nearby energy levels~\cite{Schneider2018LocalEffect}.

\section{Compatibility of the scheme with single-qubit gates}

Whereas the ability to extend the coherence lifetime of a quantum system is desirable, we would also like to manipulate this system while protecting it from the environment. For this reason, in this section we will investigate if the discussed protection scheme is compatible with the application of coherent gates. Here we will restrict to the case of single-qubit gates. As examples we will consider a $\pi$-pulse $U_{X_\pi}$ and a Hadamard gate $U_H$. 
Denoting a general rotation by an angle of $\theta$ around the $A$ axis of the Bloch sphere by $R_A(\theta) = \exp(-i \theta A/2)$, the latter gates correspond to $U_{X_\pi} =  R_X(\pi)$ and  $U_H = R_X(\pi) R_Y(\pi/2)$.
For this analysis we will consider a TLS in a slowly fluctuating environment described by RTN fluctuations of its transition frequency and we will neglect energy relaxation. 

We apply these gates via coherent pulses on the TLS and for simplicity we consider  square pulses. 
A general rotation $R_A(\theta)$ will correspond to a pulse of the form $H_{\rm drive}(t) = \Omega A [\Theta(t-t_0) - \Theta(t-t_f)]$, with $\Omega$ the Rabi frequency of the drive, $\Theta(t)$ the Heaviside function, $t_0$ and $t_f$ the times at which we turn on and turn off the pulse respectively,
and where the desired rotation angle is the pulse area, i.e., $\theta = \Omega\, t_{\rm drive}$, with $t_{\rm drive} = t_f - t_0$ the duration of the pulse. 
We evaluate the fidelity of the gate derived from these pulses in the presence of noise with and without continuous protection by simulating quantum process tomography (QPT)~\cite{Nielsen2010QuantumInformation}. 

A quantum process is mathematically represented by a completely positive linear map $\mathcal{E}$.
The goal of QPT is to estimate $\mathcal{E}$ from experimental measurements.
A quantum process acting on an arbitrary quantum state $\rho$ can be cast in the Kraus representation
\begin{align}
    \mathcal{E}(\rho) = \sum_k E_k \rho E^\dagger_k,
\end{align}
with $E_k$ the so-called Kraus operators. We can choose an arbitrary operator basis $\{ \tilde{E}_k \}$ in which to represent the Kraus operators, i.e., $E_k = \sum_i e_{ki} \tilde{E}_i$. For the case of a single TLS a natural choice is the Pauli basis $\{I, X, Y, Z \}$ which we adopt here. Therefore, the above relation can be re-written as 
\begin{align}\label{eq:chi-matrix}
    \mathcal{E}(\rho) = \sum_{ij} \chi_{ij} \tilde{E}_i \rho \tilde{E}^\dagger_j,
\end{align}
with $\chi_{ij} = \sum_k e_{ki} e_{kj}^*$ the so-called \textit{process} or \textit{chi}-matrix. Once we choose the operator basis the process matrix completely characterizes the process. In order to numerically determine the chi-matrix, we prepare the 
set of initial states
$\{\ket{1},\, \ket{2},\, (\ket{1} + \ket{2} )/\sqrt{2},\, (\ket{1} +i \ket{2} )/\sqrt{2}\}$. Each of these initial states is subject to the combined action of the pulses, RTN and correction drives. The corresponding output states are reconstructed via quantum state tomography upon gathering a considerable large measurement statistics, in our case $10^4$ independent measurements for each of the above states. Finally, the chi-matrix is obtained by solving equation~\eqref{eq:chi-matrix}. 

\begin{figure}
\centering
\includegraphics[width=1.0\linewidth]{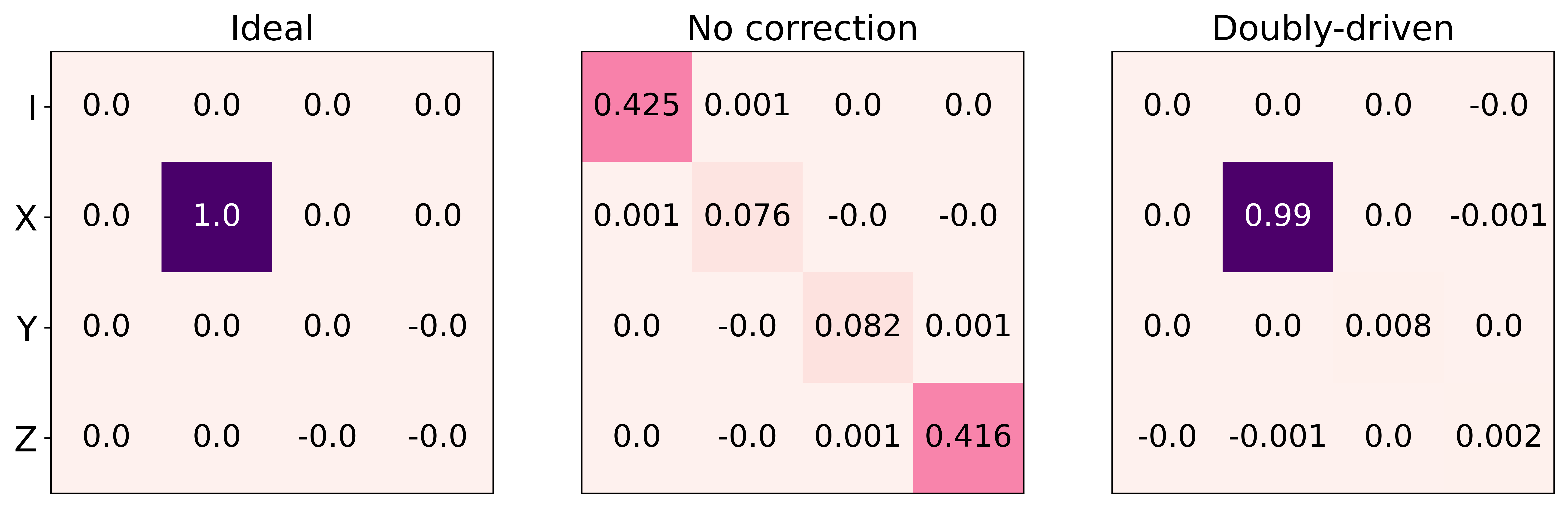}
\includegraphics[width=1.0\linewidth]{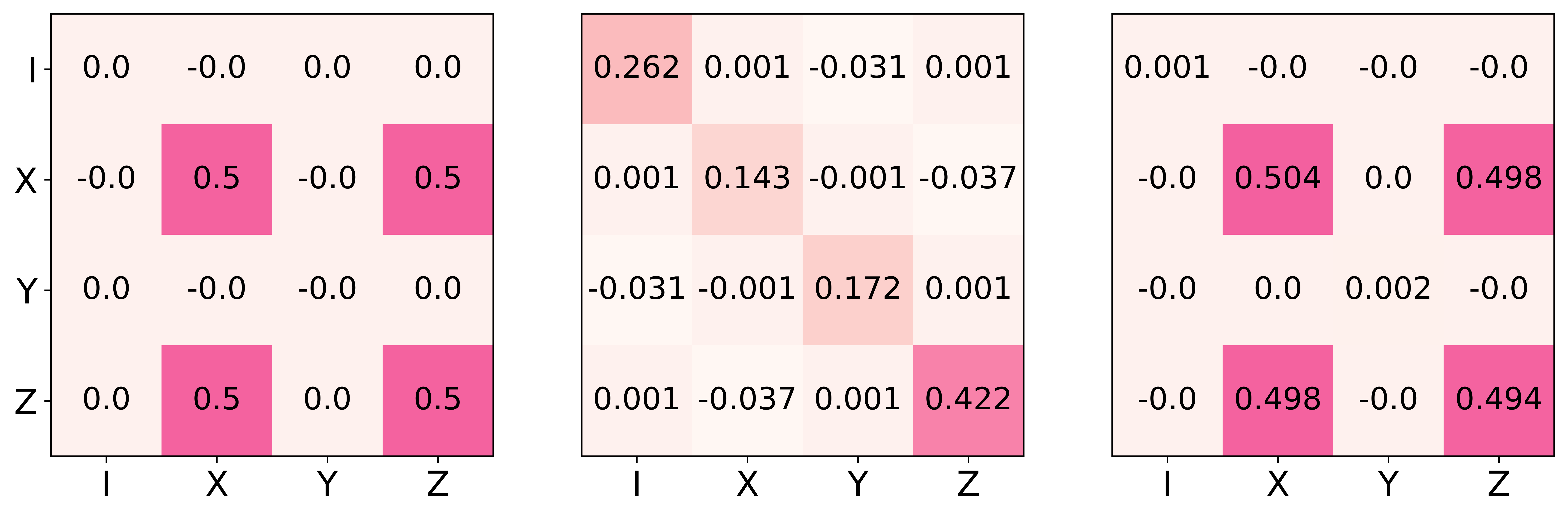}
\caption{Quantum process matrices for a $X_\pi$-pulse (top row) and a Hadamard gate (bottom row) in the presence of slow RTN with $\xi = 8/\tau$ and $\chi = 1/ \tau$. 
For both processes the left column represents the ideal gate in the absence of noise, the middle column corresponds to the gate applied in the presence of the noise but without protection, and the right column represents the gate applied in the presence of the noise while simultaneously protecting the TLS with $s=80$ using the doubly-driven scheme with detuning $\Delta =4000/\tau$.  }
\label{fig:qpt}
\end{figure}

Scaling time in units of $1/\tau$, the correlation time of the environment,  the reconstructed chi-matrix of the resulting gate process where we have chosen: a RTN with amplitude $\xi = 8/\tau$, jump rate $\chi = 1/\tau$, an emitter with sensitivity $s=80$ and drives' detuning $\Delta = 4000/ \tau$ are shown in Fig.~\ref{fig:qpt}. 
The top row corresponds to $U_{X_\pi}$ and the bottom row to $U_H$. For both cases, the left column corresponds to the ideal process matrix, i.e., applying the pulses realizing the gates to the TLS in the absence of the RTN. The middle column corresponds to the process matrix when we apply the pulses in the presence of the RTN without the protection active. For the case of a slowly drifting environment, the gate pulses are mostly off-resonant with the TLS which results in rather low process fidelities $F = 0.08$ for $U_{X_\pi}$ and $F = 0.25$ for $U_H$. Finally, the right column corresponds to the gates realized through the pulses in the presence of the RTN and the doubly-driven protection scheme. It is straightforward to see that we almost recover the ideal case process matrices. This is confirmed by the very high achieved process fidelity $F = 0.99$ for both $U_{X_\pi}$ and $U_H$.

\section*{Conclusions}

In this work we have shown that the continuous dynamical decoupling schemes introduced in Refs.~\cite{Yavuz2013SuppressionShift,Finkelstein2021ContinuousDephasing} also allow to mitigate  pure dephasing noise induced on a single TLS by a fluctuating environment with a finite correlation time. In a simple classical phase randomization picture, the fluctuations of the environment induce fluctuations in the TLS frequency which lead to the average loss of its coherence.
The availability of an additional energy level with an enhanced sensitivity to the environment
allows the emitter system to dynamically compensate for these non-Markovian temporal fluctuations via the ac-Stark shift thus greatly enhancing the TLS coherence time.


Through numerical simulations we have established that this correction scheme is independent of the Gaussian or non-Gaussian nature of the fluctuating environment. Furthermore, it is not limited by the correlation time of the latter. We have shown qualitatively, for the studied noise models, that for the protection to work the drives' detuning needs to overcome the inverse correlation time of the environment. 

Further, we have shown that the protection scheme is compatible with the application of single-qubit gates realized by pulses of finite duration. This is a key element in order for the protection scheme to be useful beyond quantum memories, for instance, for quantum information processing tasks.  Investigating a particular physical implementation of the correction scheme is out of the scope of the present work. Nevertheless, 
this scheme is ideally suitable to spin systems sensitive to fluctuating external fields. 

Finally, extrapolating from \cite{Finkelstein2021ContinuousDephasing}, this scheme  should be capable of operating in parallel on an ensemble of emitters to provide protection against both spatial and temporal noise.

\section{Acknowledgements}

The authors acknowledge support from the Okinawa Institute of Science and Technology Graduate University and in particular the
super-computing
facilities provided by the Scientific Computing and Data
Analysis section of the Research Support Division of OIST.

\bibliography{references-spinsf.bib}
\end{document}